\documentclass{ws-rv10x7}
\usepackage{ws-rv-van}     
\usepackage{ws-rv-thm}     
\usepackage{subfigure}
\makeindex
\usepackage{ws-index}      
\begin{document}

\renewcommand{\thechapter}{27}

\chapter[Glassy features and complex dynamics in ecological systems]{Glassy features and complex dynamics in ecological systems\\ \label{Ch 27}}

\author[A. Altieri]{A. Altieri\footnote{ada.altieri@u-paris.fr}}

\address{Laboratoire Matière et Systèmes Complexes (MSC), Université Paris Cité\\
CNRS, 75013 Paris, France }

\vspace{0.3cm}

\begin{abstract}
In this report, I will review some of the most used models in theoretical ecology along with appealing reformulations and recent results in terms of diversity, stability and functioning of large well-mixed ecological communities.
\end{abstract}

\body

\section{Introduction}
\label{sec1}

Emergent properties of many-species ecological communities have a variety of applications: for example, the activity of the gut microbiota is believed to be crucial for human health; sustaining natural diversity is essential for services such as food supply, pollination and climate regulation. There is growing awareness that human activity is causing irreversible species extinction and ecosystem simplifications, generally considered a \emph{global biodiversity crisis}. The Earth Microbiome Project\footnote{https://earthmicrobiome.org/} and the Human Microbiome Project\footnote{\url{https://www.hmpdacc.org/}} are designed in this direction aiming to identify and characterize all diverse microorganisms and their relationship to ecological stability and disease development.

The incredible biodiversity that characterizes natural ecosystems has attracted ecologists for long time but more recently has started gathering interest also among theoretical physicists.
From a theoretical perspective, modeling the interactions between many different components – from bacteria in a microbial community to plant-pollinator impact in a forest to starling murmurations – can become extremely complicated. 
A single, well-established theory allowing one to bridge the gap between empirical data made available from an enormous number of controlled experiments and more sophisticated techniques is nevertheless still missing.
In addition to the need for a general criterion that would enable to discriminate between \emph{niche theory} -- for which each niche is occupied by a single species according to the competitive exclusion principle \cite{Hardin1960} -- and \emph{neutral models} -- in which differences are only attributed to stochasticity -- other crucial questions come to the stage and play an increasingly key role: i) relaxation either to a single fixed point or a multiple fixed point regime; ii) definition of ecosystem diversity, \emph{i.e.} the number of surviving species; iii) fluctuation and functional response typical behavior under the effect of external perturbations; iv) investigation of the interplay between stochastic and deterministic processes and how community diversity and variability are related to them; v) emergence of possible chaotic dynamics and limiting cycles to be experimentally measured.

In this chapter, we aim to present different statistical physics frameworks that rely on advanced spin-glass techniques, for which Giorgio Parisi has been a pioneer as well as a beacon outlining the right direction in a multitude of complex scenarios.

\subsection{More is Different}


Theory has long predicted that large complex systems are intrinsically unstable \cite{May1972, May1976}, which is a long-standing puzzle given the complexity observed in Nature. In the last years, there is nevertheless a growing interest in systems composed of an enormous number of species interacting in myriad ways in very complex environments. Such systems can thus be rephrased through the prism of statistical physics using sophisticated concepts and powerful methods in this direction \cite{Faust2012, Fisher2013, Kessler2015, Bunin2017, Altieri2019, Servan2018, Marsland2020, pearce2020stabilization, wu2021understanding}. 
In a bottom-up approach, the detailed structure of individual interactions and how such coefficients scale with the system size is unknown since particularly difficult to infer in diversity-rich ecosystems. 
Hence, to tackle the staggering complexity of large ecological communities, one can follow a long tradition rooted in Robert May's seminal works \cite{May1972, May1976} and assume the interaction matrix to be random.  May considered a \emph{community matrix} $H$ of size $S \times S$, $S$ being the total number of species in the pool and $H_{ij}$ standing for the effect of species $j$ on $i$ around a feasible fixed point. In this picture, the self-regulation term corresponding to diagonal elements is fixed to $-1$, whereas off-diagonal elements are drawn from a random distribution with zero mean and variance $\sigma^2$ -- sometimes referred to as heterogeneity parameter -- with associated probability $C$. According to May's conjecture, if $\sigma \sqrt{S C} >1$ the system is inevitably unstable under infinitesimally small perturbations and cannot persist. Hence, as a system becomes more diverse (controlled by the number of species $S$ in the pool), more connected (in terms of the connectivity $C$), and strongly interacting (tuned by $\sigma$), a transition to instability occurs with a probability of persisting close to zero.
In the large $S$ limit, random matrix theory comes into play claiming that the eigenvalues of the community (or Jacobian) matrix must be contained inside a circle of radius $\sigma \sqrt{SC}$ in the complex plane. Therefore, the system's stability is conditional on the fact that the resulting circle is located in the left half-plane with all eigenvalues having negative real parts. 

To provide general criteria that could encompass all diversified cases, one can then play with the interaction matrix by changing the strength and mutual sign. A suitable reshuffling of local interactions clearly raises a number of questions on how different combinations of them affect the stability of the overall community and what would be a good trade-off (weak/strong, mutualistic/competitive) to avoid, for instance, destabilization of a prey-predator chain if weak interactions are preponderant \cite{Allesina2012}.


\section{High-dimensional MacArthur model at the edge of stability}

In the following, we shall focus on mathematical models that offer a suitable platform to understand ecosystems' behavior: giving some input information, predictions on species survival, responses to external perturbations, and the emergence of robust structures can be extracted as an output. We will start with a very influential one, the MacArthur resource-consumer model, originally designed to shape competition among $S$ different species for $N$ non-interacting resources \cite{Macarthur1970}.
Notably, if the dynamics describing resource evolution is much faster than the populations' one, the former can be integrated out leading to the generalized Lotka-Volterra equations\cite{Lotka1920, Volterra1927}. The random Lotka-Volterra model will thus represent the second core of this chapter, through which we will figure out how to overcome certain inherent limitations of such a resource-consumer model.

By taking advantage of the definition of self-averaging quantities, MacArthur's model has been recently reformulated as a problem of statistical physics of disordered systems and then solved analytically in the limit of an infinite number of species and resources \cite{Tikhonov2017}. We will especially use it to probe several underlying connections between the phenomenology of jamming \cite{Altieri2019book} and criticality in large ecosystems.

The dynamics of the model is defined by linear differential equations for $n_\mu$ individuals, where the index $\mu=1,..., S$ denotes the different species: 
\begin{equation}
\frac{d n_\mu}{dt} \propto n_\mu \Delta_\mu \ ,
\label{dn-mu}
\end{equation}
and $\Delta_\mu$ is the \emph{resource surplus}. As far as one is concerned with equilibrium, the proportionality factor in the dynamical equation above can be safely neglected.  The equilibrium condition from Eq. (\ref{dn-mu}) leads to two possibilities: i) $n_\mu >0$ $\&$ $\Delta_\mu=0$ (survival); ii)  $n_\mu =0$ $\&$ $\Delta_\mu<0$ (extinction)\footnote{The case $\Delta_\mu >0$ is actually forbidden by the model definition.}. 
The variables $\Delta_\mu$ depend then on the availabilities of resources $h_i$ (with $i=1,..., N$) and the \emph{metabolic strategies}, $\sigma_{\mu i}$'s, by which species demand and possibly meet their requirement $\chi_\mu$:\begin{equation}
\Delta_\mu=\sum_{i=1}^{N} \sigma_{\mu i} h_i -\chi_\mu   \ .
\label{Delta}
\end{equation}
For each species $\mu$, the metabolic strategy represents a random binary vector whose components $\sigma_{\mu i}$ are extracted from a distribution that takes values $1$ and $0$ with probabilities $p$ and $1-p$ respectively. The parameter $p$ determines whether the species in the ecosystem are either specialists ($p \ll 1$), each requiring a small number of well-defined metabolites necessary for their survival, or generalists ($p \sim 1$), meaning that many different metabolites can be appropriate for their needs. 
In turn, individuals $n_\mu$ depend on the availability of resources, $h_i$, according to a feedback loop mechanism, which is essentially modulated by the efficiencies through which species exploit resources. 
By defining a total demand, $T_i=\sum_\mu n_\mu \sigma_{\mu i}$, the availabilities $h_i$ can simply be expressed as a decreasing function of it. For instance, one can consider $h_i= \frac{R_i}{\sum_\mu n_{\mu} \sigma_{\mu i}}$
where $R_i$ is the resource surplus whose average is constant whereas its variance, $\delta R^2$, can fluctuate and be used to reproduce the resulting phase diagram.


Over the years several mechanisms have been put forward to explain the fact that complex -- and in particular living -- systems tend to be poised at the edge of stability: edge of chaos \cite{Kauffman1991}, self-organized criticality \cite{Bak2013nature}, self-organized instability, scale-free behavior, etc. Here we propose an example that leverages on an alternative principle\cite{Altieri2019}. It is based on recasting the MacArthur model in terms of a Constraint Satisfaction Problem (CSP). Hence, in analogy with a standard CSP, above the hyperplane $\vec{h} \cdot \vec{\sigma}_\mu$ species are able to survive and multiply; conversely, if $\vec{h} \cdot \vec{\sigma_{\mu}} < \chi_\mu$, the sustainability of the species' pool is no longer guaranteed.
All $\vec{h}$ such that $\vec{h} \cdot \vec{\sigma_\mu} < \chi_\mu$ define the so-called \emph{unsustainable region}, for each species $\mu$.
One can now re-express the requirement $\chi_\mu$ via a random variable \emph{i.e.} $\chi_\mu=\sum_i \sigma_{\mu i} +\epsilon x_\mu$ \cite{Tikhonov2017}, where the parameter $\epsilon$ plays the role of an infinitesimal cost scatter and $x_\mu$ is a zero-mean and unit-variance Gaussian variable. It has been shown that, in the $\epsilon \rightarrow 0$ limit, the model undergoes a phase transition between two qualitatively different regimes: i) a \emph{shielded phase}; ii) a \emph{vulnerable phase} \cite{Tikhonov2017}.
In the shielded phase, $\mathcal{S}$, a collective behavior emerges with no influence of external conditions. If the availabilities are set to one in such a way that neither specialists nor generalists are favored, and a sufficiently small perturbation is applied to the system, a feedback mechanism between $h_i$ and $n_\mu$ contributes to adjusting mutual species' abundance and to keeping the availabilities almost unchanged, $\forall i$. 
The situation is quite different in the \emph{vulnerable phase}, $\mathcal{V}$, where species cannot self-sustain and turn out to be strongly affected by changes and improvements in the immediate environment.


To characterize the stability of a general competing system against perturbations in a more rigorous way, one can introduce a Lyapunov function and compute the density of fluctuations in the two phases.
The positive or vanishing behavior of such a function, together with its time derivative, provide information on whether the equilibrium is unstable, locally asymptotically stable, or globally asymptotically stable. 
In this specific case, the Lyapunov function reads
\begin{equation}
F(\lbrace n_\mu \rbrace)=\sum_{i} R_i \log \left(\sum_\mu n_\mu \sigma_{\mu i} \right) -\sum_\mu n_\mu \chi_\mu \ ,
\label{lyapunov}
\end{equation}
which is bounded from above, hence guaranteeing that an equilibrium always exists. By differentiating Eq. (\ref{lyapunov}) to the second order, one eventually obtain
\begin{equation}
\frac{d ^2 F}{d n_\mu d n_\nu}=-\sum_{i} \sigma_{\mu i} \sigma_{\nu i} \frac{R_i}{(\sum_{\rho} n_{\rho} \sigma_{\rho i} )^2}=-\sum_{i} \sigma_{\mu i} \sigma_{\nu i} \left( \frac{h_i^2}{R_i} \right) \ .
\label{Hessian_n}
\end{equation}
In the $\mathcal{S}$ phase, \emph{i.e.} for $h_i \simeq 1$, this expression leads to a modified Wishart matrix whose eigenvalue distribution is defined by a Marchenko-Pastur law \cite{MarchenkoP} in the limit of a large number of species and resources. Accordingly, the resulting spectral density reads:
\begin{equation}
\rho(\lambda)= \frac{1}{2 \pi} \frac{\sqrt{(\lambda-\lambda_{-})(\lambda_{+}-\lambda)}}{\lambda} \ ,
\end{equation}
where the upper and lower edges of the spectrum are $\lambda_{\pm}= (\sqrt{[1]}\pm 1)^2$. The quantity $[1]$ denotes the fraction of active species at criticality or, borrowing the Constraint Satisfaction Problem (CSP) jargon, the fraction of \emph{satiated constraints} for which $\Delta_\mu=0$. 
In analogy with the so-called SAT/UNSAT transition, we can associate the $\mathcal{V}$ phase to a \emph{hypostatic regime}, with a smaller number of saturated constraints with respect to the total number of variables \cite{Wyart2005, Franz2016}: this case corresponds to a gapped spectral density without any signature of an emerging criticality.
Conversely, the $\mathcal{S}$ phase would correspond to an \emph{isostatic regime} -- where the number of vanishing constraints equals the overall space dimension, and a gapless spectrum for the distribution of eigenvalues appears. Because the lower edge of the spectrum $\lambda_{-} \rightarrow 0$ tends to zero upon approaching the $\mathcal{V}/\mathcal{S}$ transition line, the eigenvalue density contribution in the $\mathcal{S}$ phase becomes:\begin{equation}
\rho(\lambda) \sim \sqrt{\left(4-\lambda\right)/{\lambda}} \ .
\end{equation}
A vanishing lower edge is in turn related to the appearance of a zero mode in the Hessian matrix of the replicated free energy (so-called \emph{replicon eigenvalue}): this translates into a diverging spin-glass susceptibility \cite{MPV, DeDominicis2006} as further evidence of being close to a critical point. A large response function can be interpreted as the fact that -- rather than being governed by a single leader -- the system tends to self-organize and respond collectively to external perturbations \cite{Mora2011}. 

It is worth noticing that since the Lyapunov function in Eq. (\ref{lyapunov}) is convex everywhere, a replica-symmetry-broken regime cannot occur. The most likely scenario taking place here is akin to the phenomenology of a \emph{random linear programming} problem\cite{Franz2016}. Even though replica symmetry continues to hold, a marginally stable regime takes place for some specific values of the control parameters.

The advantage of introducing a high-dimensional version of the MacArthur model is that it provides an appealing and easily-defined reference model albeit, in its current form, lends itself to describing only competitive interactions. To suitably address a wider spectrum of ecological scenarios, the random version of the Lotka-Volterra model will be presented in the following accounting either for the competitive or cooperative case.

\section{The generalized random Lotka-Volterra model}

A wide range of phenomena in population dynamics, including predation, mutualism, and resource-consumer interactions, can be reasonably well captured by a much simpler reference model: the disordered Lotka-Volterra model whose typical features are shown off by tuning a few control (universal) parameters. Moreover, it not only reproduces phenomenologically multiple facets of well-mixed ecosystems \cite{Barbier2018} but also turns out to be of great interest in interdisciplinary domains such as genetics, epidemiology\cite{holt1985infectious}, and evolutionary game theory \cite{Galla2013, Sanders2018} up to the modelization of complex financial markets\cite{Sprott2004, Moran2019}. 
The Lotka-Volterra equations describe the evolution of $S$ species subject to random interactions $\alpha_{ij}$\cite{Kessler2015,Bunin2017}:
\begin{equation}
    \frac{d N_i}{dt}= N_i \left[1-N_i -\sum_{j, (j \neq i)} \alpha_{ij} N_j \right] +\sqrt{N_i} \eta_i(t) +\lambda_i \ ,
    \label{dynamical_eqT}
\end{equation}
where $N_i(t)$ is the relative abundance of species $i$ (with $i=1,...,S$) at time $t$ meaning that the population is normalized with respect to the total number of individuals $N_\text{ind}$ that would be present in the absence of interaction.
The elements of the random matrix $\alpha_{ij}$ are independent and identically distributed with mean $\langle \alpha_{ij}\rangle =\mu/S$, variance $\langle \alpha_{ij}^2\rangle_c=\sigma^2/S$ and $ \langle \alpha_{ij} \alpha_{ji} \rangle_c=\gamma \langle \alpha_{ij}^2\rangle_c$, where the subscript $_c$ stands for the connected part of the correlation. The parameter $\gamma$ ranges from $-1$ (completely antisymmetric case to which prey-predator interactions belong) to $1$ (fully symmetric, for which a Lyapunov function can be safely defined).

The demographic noise contribution, accounting for deaths, births, and other unpredictable events, is modelled by $\eta_i(t)$, a Gaussian variable with zero mean and variance $\langle \eta_i(t)\eta_j(t')\rangle=2T \delta_{ij} \delta(t-t')$, whose amplitude $T$ is inversely proportional to the total number of individuals $N_\text{ind}$.
Such a multiplicative noise term allows us to investigate the effect of demographic stochasticity in a continuous setting \cite{Domokos2004discrete, Rogers2012, weissmann_simulation_2018}: the larger the global population, the smaller the strength $T$ of the demographic noise. Then, to guarantee the probability distribution to be integrable at small abundances, we need to introduce a small but finite immigration rate, which will be assumed to be constant over species, \emph{i.e.} $\lambda_i=\lambda$. This is a smart way to avoid an absorbing boundary in $N_i=0$ due to the introduction of a finite demographic noise, which in turn would push a finite fraction of species to zero\footnote{With no demographic noise and no immigration, a similar model was proposed in the nineties by Biscari and Parisi \cite{Biscari1995} and analyzed by studying the stability of the replica symmetric solution (single fixed point regime).}.

In the case of random symmetric interactions, the stochastic process induced by Eq. (\ref{dynamical_eqT}) admits an equilibrium-like stationary distribution \cite{Biroli2018_eco, Altieri2021} with associated Hamiltonian:
\begin{equation}
    H=  {{-}} \sum_i  \left(N_i-\frac{N_i^2}{2}\right)+\sum_{i<j}\alpha_{ij}N_iN_j+ \sum_i [T\ln N_i - \ln \theta(N_i-\lambda)] \  .
    \label{Hamiltonian_0}
\end{equation}
The before-last term is due to the demographic noise\footnote{The parameter $T$ plays the role of the temperature in a statistical mechanics setting. The mapping can be easily established by writing the corresponding Fokker-Planck equation with a white Gaussian noise.} whereas the counterbalancing role of the immigration is formally modelled by is the Heaviside function $\theta(x)$, which corresponds to imposing a reflecting wall at $N_i=\lambda$.

\subsection{Glassy phases and out-of-equilibrium dynamics}

Adding a finite demographic noise not only allows us to get a more general picture but also to properly characterize the resulting phase diagram -- see Fig. 2 -- connecting peculiar properties of each regime to the ones of equilibria.

\begin{figure}[h]
\center
\includegraphics[scale=0.36]{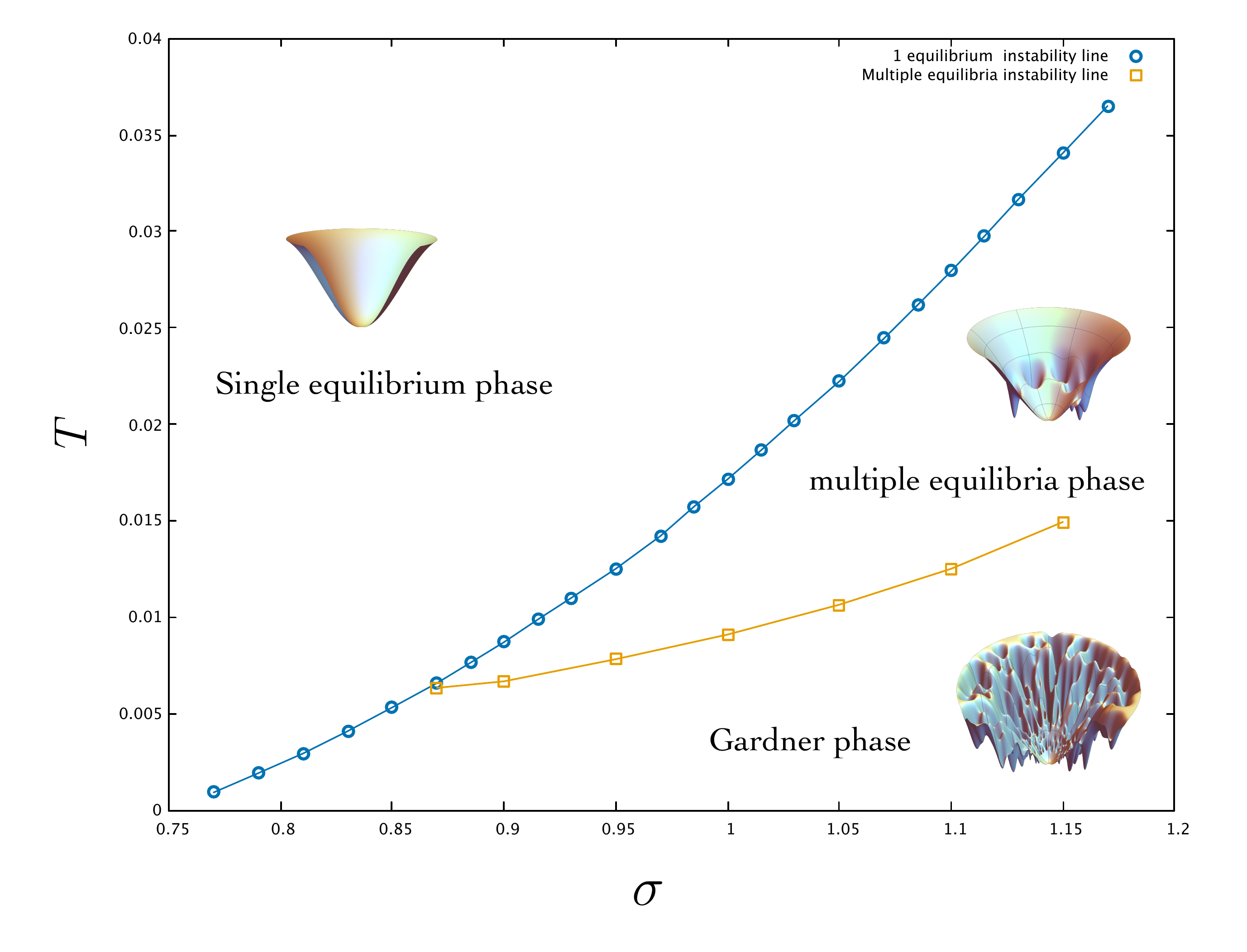}
\caption{Phase diagram showing how the variation of the demographic noise strength, $T$, and the heterogeneity of interactions, $\sigma$, can lead to three different phases. 
In particular: i) a single equilibrium phase where the configurational landscape is purely convex; ii) a multiple equilibria regime, which is characterized by a $1$RSB stable solution and an exponential number of locally stable equilibria; iii) a \emph{Gardner phase}, which turns out to be associated with a hierarchical organization of the equilibria in the free-energy landscape. Figure taken from \cite{Altieri2021}.}
\end{figure}

\begin{figure}[h]
\centering
\begin{minipage}{.5\textwidth}
  \centering
  \includegraphics[width=0.87\linewidth]{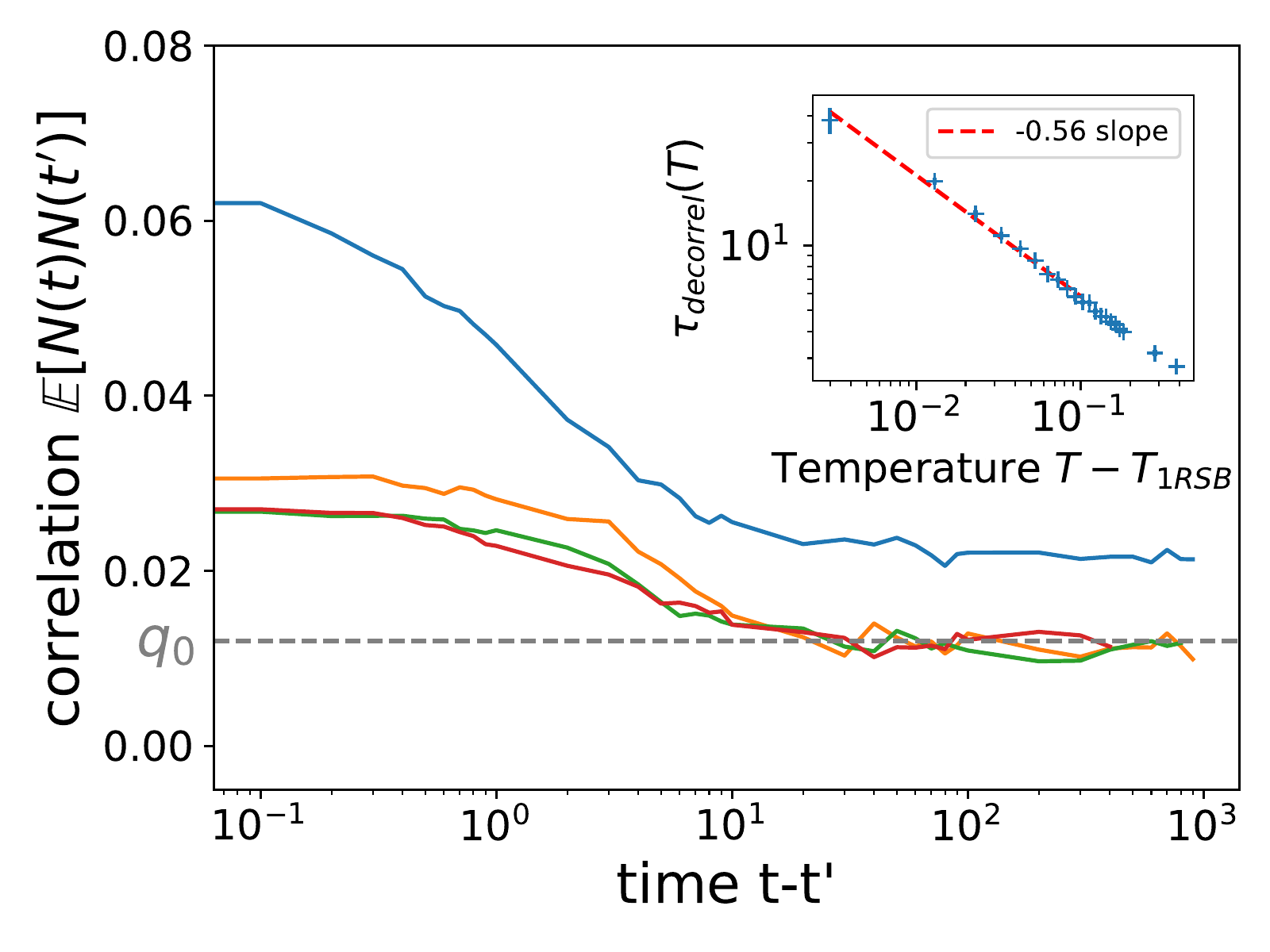}
  \label{fig:test1}
\end{minipage}%
\begin{minipage}{.5\textwidth}
  \centering
  \includegraphics[width=0.85\linewidth]{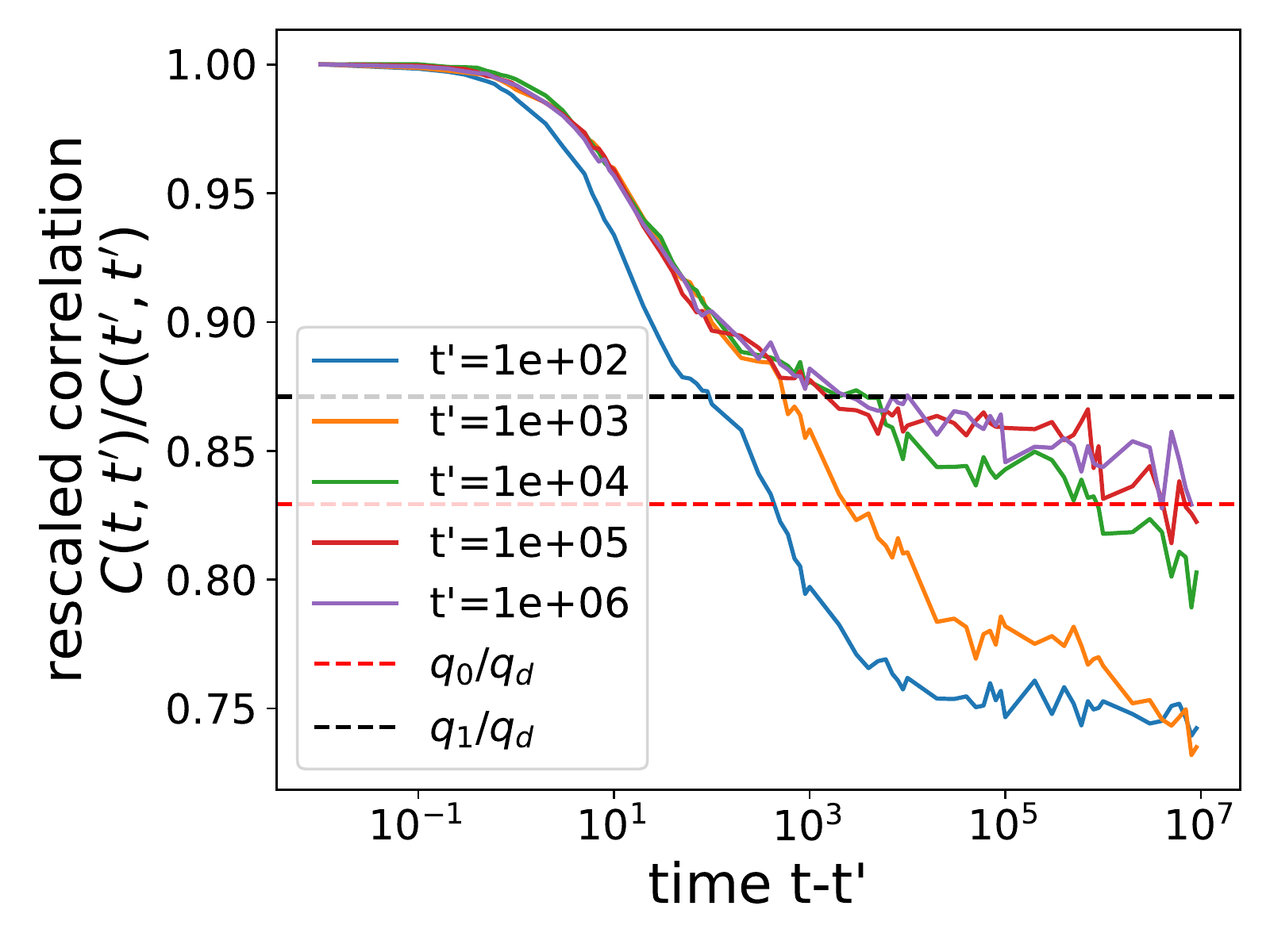}
  \label{fig:test2}
\end{minipage}
\caption{Numerical simulations based on DMFT. Two-time correlator $C(t,t')$  in the single-equilibrium phase (RS, on the left) compared with the same correlator in the multiple equilibria phase (1RSB, on the right) plotted for different $t'$ and $S=500$. The dashed lines correspond to the values of the overlap parameters, which are obtained by the replica method. The inset in the left plot highlights a divergence in the decorrelation time as $T \rightarrow T_\text{$1$RSB}$, the critical temperature associated with an instability of the RS solution. Figures taken from \cite{Altieri2021}.}
\end{figure}
Then one may wonder how all these outcomes are expected to change when asymmetric interactions are also taken into account and which strategy proves to be the most appropriate in this case.
Non-symmetric interactions strongly complicate the analysis since they correspond to plugging non-conservative forces in the dynamics thus violating the Fluctuation-Dissipation theorem (FDT) and bringing the system out of equilibrium. Since it is no longer possible to define a Hamiltonian to be minimized and analyzed in terms of harmonic fluctuations around each of the minima,
the cavity \cite{MPV, Mezard2009} and Dynamical Mean-Field Theory \cite{Roy2019numerical, Altieri2020dyn} formalisms come into play. The last method, in particular, allows us to map a multi-variable problem into a single-body stochastic formalism, which eventually involves time-delayed friction and colored noise whose features have to be determined self-consistently. In other words, the two-time correlation $C(t,t')$ and response $R(t,t')$ functions are fixed in a self-consistent way given the probability distribution associated with the stochastic process and the distribution of random interactions. 

A similar analysis, as illustrated for the symmetric case in Fig. (2), can be performed.  Without demographic fluctuations, increasing the variability of the interactions $\sigma$ would destabilize the single-fixed-point regime and eventually result in chaotic phases as for neural networks and spin-glass models in the presence of asymmetric couplings.
The introduction of a positive immigration rate would lead to the stabilization of chaotic dynamics -- with an indefinitely long lifetime -- corresponding to what we have referred to as \emph{multiple equilibria regime} in the purely symmetric case. However, as soon as the immigration rate is set to zero, the chaotic regime is no longer stable \cite{Roy2020, pearce2020stabilization}, replaced by slower and slower dynamics (\emph{aging}).

\subsection{Non-logistic growth functions and pseudo-gap distributions}

The Lotka-Volterra equations analyzed in the large-$S$ limit thus far allow us to get analytical advances in a very broad class of problems. In particular, by slightly modifying the dynamical Eq. (\ref{dynamical_eqT}) through the introduction of a higher-order one-species potential, one can also investigate the so-called \emph{Allee effect}\cite{Allee1926}, which describes a positive correlation between mean individual fitness (or per-capita growth rate) and population density over some finite interval\cite{Gascoigne2004, Kramer2018}. This positive feedback loop mechanism, which inherited the name from the famous zoologist Allee, essentially relies on the observation that in many species under-crowding, and not only competition, contributes to limiting population growth.  The Allee effect is called \emph{strong} if there exists an initial population threshold in the sense that the species pool needs a sufficiently large initial population to avoid extinction, whereas it is denoted as \emph{weak} if no threshold exists. Even in this second case, intra-specific cooperation leads to an initial increase in the growth rate as population increases (see Fig. (3)).

\begin{figure}[h]
\center
\includegraphics[scale=0.28]{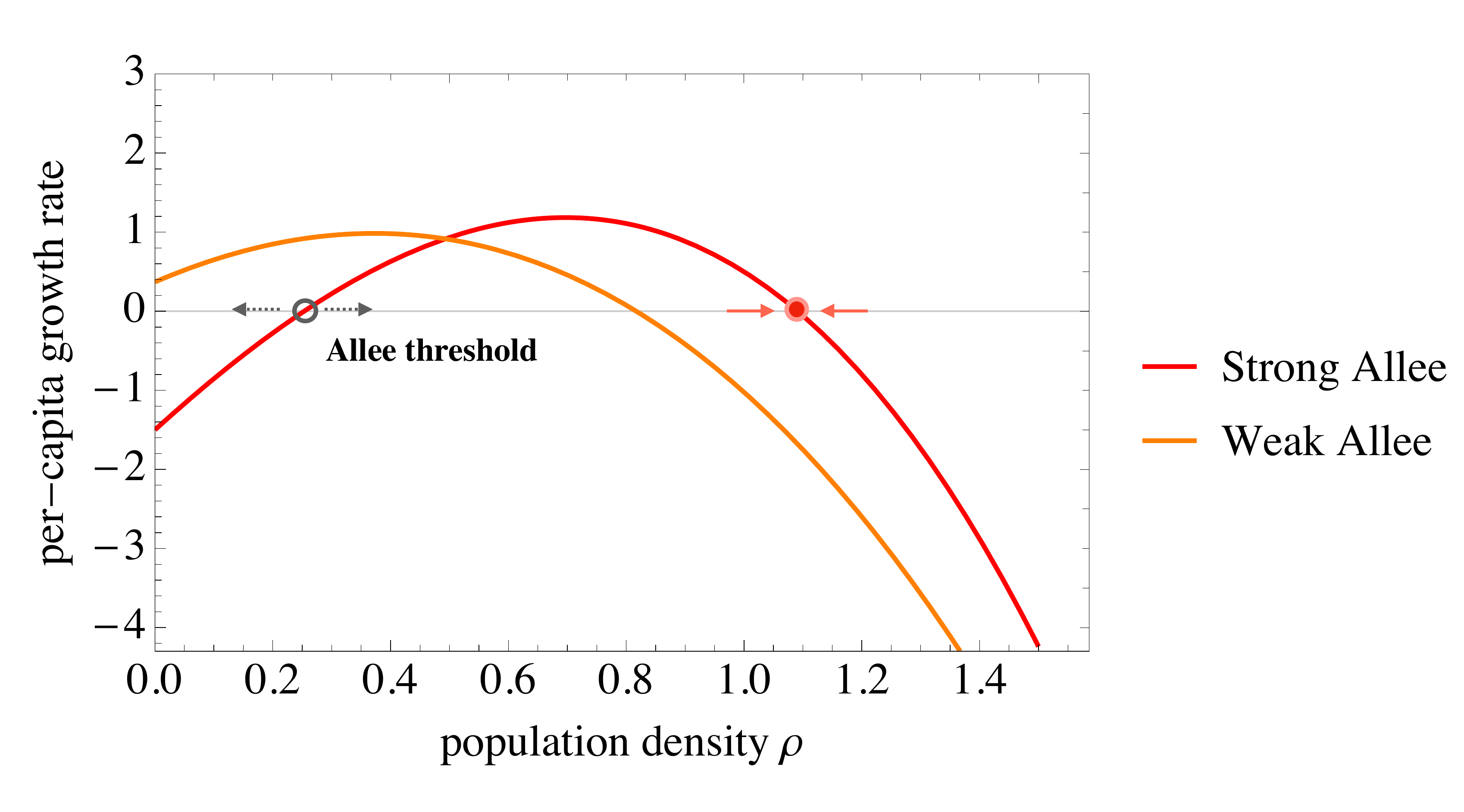}
\caption{Sketch of the strong Allee effect (in red) compared to a weak Allee effect (in orange). In the former, the finite threshold corresponds to an unstable fixed point (empty black circle); in the latter, no threshold in the population exists.}
\end{figure}
In the same spirit as before, one can take advantage of thermodynamic analysis and shed light on the resulting phase diagram by tuning the strength of random interactions and the demographic noise. Remarkable differences emerge with respect to the Lotka-Volterra logistic-growth case \cite{Altieri2022}. First, the number of states below the critical transition line is no longer exponential in the system size nor separated by extensive barriers, exactly as it would happen in equilibrium states of mean-field spin glasses (\emph{i.e.} the Sherrington-Kirkpatrick model \cite{MPV}). Furthermore, as soon as one considers a non-linear functional response of the species abundances, a pseudo-gap distribution in the local curvatures of the single-species effective potential appears\footnote{With the exponent $\alpha \ge 1$.}, $P(V^{''}_\text{eff}(N^{*})) \sim \vert V^{''}_\text{eff}(N^{*}) \vert^{\alpha}$, as a clear signature of a marginal low-demographic noise (low-temperature) phase\cite{Altieri2022}.
This outcome nicely generalizes the pseudo-gap distribution that was found for instantaneous local fields in mean-field spin glasses -- and was obtained before only in the case of discrete degrees of freedom\cite{Palmer1979} -- to a complex ecological model.

\section{Conclusions and perspectives}

Along the different sections of this short report, I have mostly discussed  analytical outcomes made possible by the use of mean-field limits.
These pages are therefore intended as a tribute to Giorgio Parisi, a way to thank him for the innovative and insightful techniques I have been learning over the years, and that have been successfully applied to such diverse and interdisciplinary contexts.

As for future research, an interesting direction would be the investigation of spatially extended models either in a completely-connected topology where multiple patches (locations in space) are coupled by diffusion or in a sparse network with finite connectivity on each site. On the one hand, this metapopulation scenario, as originally proposed by Levins\cite{Hanski1998, Hanski2000, Etienne2002}, would allow us for a more tangible comparison with real data, starting for instance with populations of small mammals and insects\cite{Elmhagen2001}; on the other hand, new appealing phenomena -- such as pattern formation, traveling waves and activity fronts \cite{Curatolo2020, Manna2021} -- are expected to appear. A rigorous theoretical analysis with an increasingly large number of species and, possibly, not only pairwise interactions is still missing.

A parallel line of research would concern an in-depth analysis of the role of different kinds of fluctuations -- demographic and environmental ones that might violate Detailed Balance – and their interplay with the deterministic dynamics. Such a classification will drive a better comparison with observational data, in particular for reproducing Species Abundance Distributions (SAD) of large ecological communities as well as for achieving a deeper understanding of the formal expression of functional responses given by local perturbations. This information would be extremely useful in the attempt to recover power-law and log-normal distributions for the species abundances that have not yet been identified in models accounting only for demographic fluctuations and symmetric interactions \cite{Lorenzana2022}.

\newpage

\section*{Acknowledgments}

I thank Matthieu Barbier, Giacomo Gradenigo and Frédéric van Wijland for a critical reading of the draft.

\bibliographystyle{ws-rv-van}
\bibliography{fullBibli}

\end{document}